# BEAST DB: Grand-Canonical Database of Electrocatalyst Properties

Cooper Tezak[1], Jacob Clary[2], Sophie Gerits[1], Joshua Quinton[3], Benjamin Rich[4], Nicholas Singstock[1,5], Abdulaziz Alherz[6], Taylor Aubry[2], Struan Clark[2], Rachel Hurst[2], Mauro Del Ben[7]*, Christopher Sutton[8]*, Ravishankar Sundararaman[3,9]*, Charles Musgrave[1,4,10,11,12]*, Derek Vigil-Fowler[2]*

Affiliations:

1. Department of Chemical and Biological Engineering, University of Colorado Boulder, Boulder, CO 80309, USA
2. Materials, Chemical, and Computational Science Directorate, National Renewable Energy Laboratory, Golden, CO 80401, USA
3. Department of Physics, Applied Physics and Astronomy, Rensselaer Polytechnic Institute, Troy, NY 12180, USA
4. Department of Chemistry, University of Colorado Boulder, Boulder, CO 80309, USA
5. Department of Mechanical Engineering, University of Colorado Boulder, Boulder, CO 80309, USA
6. Department of Chemical Engineering, College of Engineering and Petroleum, Kuwait University, Safat 13060, Kuwait
7. Applied Mathematics & Computational Research Division, Lawrence Berkeley National Laboratory, Berkeley, CA 94720, USA
8. Department of Chemistry and Biochemistry, University of South Carolina, Columbia, SC, 29208, USA
9. Department of Materials Science and Engineering, Rensselaer Polytechnic Institute, Troy, NY 12180, USA
10. Materials Science and Engineering Program, University of Colorado Boulder, Boulder, CO 80309, USA
11. Renewable and Sustainable Energy Institute, University of Colorado Boulder, Boulder, CO 80309, USA
12. Department of Chemical Engineering, University of Utah, Salt Lake City, UT 84112, USA

Corresponding Emails:
mdelben@lbl.gov
cs113@mailbox.sc.edu
sundar@rpi.edu
charles.musgrave@colorado.edu
Derek.Vigil-Fowler@nrel.gov

## Abstract

We present BEAST DB, an open-source database comprised of ab initio electrochemical data computed using grand-canonical density functional theory in implicit solvent at consistent calculation parameters. The database contains over 20,000 surface calculations and covers a broad set of heterogenous catalyst materials and electrochemical reactions. Calculations were performed at self-consistent fixed potential as well as constant charge to facilitate comparisons to the computational hydrogen electrode. This article presents common use cases of the database to rationalize trends in catalyst activity, screen catalyst material spaces, understand elementary mechanistic steps, analyze electronic structure, and train machine learning models to predict higher fidelity properties. Users can interact graphically with the database by querying for individual calculations to gain granular understanding of reaction steps or by querying for an entire reaction pathway on a given material using an interactive reaction pathway tool. BEAST DB will be periodically updated, with planned future updates to include advanced electronic structure data, surface speciation studies, and greater reaction coverage.

# 1. Introduction

Electrocatalysis is a cornerstone technology that underlies numerous renewable energy and chemical processes crucial for realizing a decarbonized economy.[1] Electrocatalysts are expected to play a central role in the production of sustainable chemicals, such as the production of green hydrogen, providing both sustainable fuel, feedstocks to other chemical processes, and long-term energy storage solutions.[2] Additionally, electrocatalysts facilitate the generation of electricity by fuel cells, enhancing their efficiency and minimizing the environmental impact of converting chemical energy back into electrical energy.[3] Beyond providing energy solutions, electrocatalysis enables circularization of the chemical industry through sustainable generation of value-added chemicals. This includes the electrochemical reduction of carbon dioxide into carbon-neutral products and the production of ammonia for fertilizer from alternative nitrogen sources.[4,5]

Advances in catalyst design and performance therefore hold significant potential for reducing our reliance on fossil fuels and minimizing environmental impact. However, gaining an understanding of electrocatalytic processes is partly hindered by the wide variety of conditions (e.g. potential ranges, catalyst composition, pH) under which experiments are performed. This diversity makes it challenging to determine the individual impact of these factors on performance.

Computational approaches offer a valuable means to rationalize these choices by providing mechanistic insight and enabling identification of trends across data more quickly than experiment.[6–8] However, they also face the complexity of simulating a wide variety of experimental conditions, which can complicate such comparisons. Data science offers promise in addressing the challenges posed by the diverse conditions encountered in electrocatalysis research, providing a comprehensive framework for analysis of electrochemical performance and development of standardized protocols and methodologies. In the materials domain, the ability to readily access material properties, including relative stability, structure, and electronic structure through a simple database query has significantly benefited researchers. This has been facilitated through large public databases of consistently computed properties such as the Materials Project,[9] NOMAD,[10] OQMD,[11] AFLOW,[12] and AiiDA.[13] These resources enable researchers to develop novel approaches for discovering functional materials and use the trends within the compiled data to guide further research.

While materials databases have paved the way for accessing computed properties, catalysis databases face unique challenges in capturing the complexity of catalytic systems. In the catalysis space, multiple databases exist with different foci: machine learning DFT structural relaxations,[14] collating data on different catalysts and chemistries for gas-phase heterogeneous catalysis based on existing literature,[15] and developing scaling relations over relatively large combinatoric spaces of metallic electrocatalysts.[16] While these databases have provided many insights within their respective areas of focus, they share common characteristics that limit their ability to inform electrocatalysis studies, they; 1) either neglect the applied potential and solvent effects or treat them at the Computational Hydrogen Electrode (CHE) level,[17] 2) compute only adsorption energies or reaction energetics, and 3) use DFT as their most accurate level of theory. Additionally, only Open Catalyst Project uses a uniform set of calculation parameters for all the calculations tabulated in the database.

The compiled electrocatalyst properties tabulated within current databases to date[14,16] have used the CHE[17] to approximate the effect of the applied potential on reaction energetics. The CHE model is the most widely used approach to include the effect of potential in electrochemical processes primarily because it only requires that one apply a simple potential-dependent, post-processing correction that shifts the energetics of proton coupled electron transfer (PCET) steps to otherwise standard uncharged DFT calculations.[18] Since its conception, CHE has been successfully used to rationalize observed trends in electrochemical reactions and design electrocatalysts that facilitate desired electrochemical conversions.[19,20] While the CHE approach has been successful, it is inherently limited in modeling the complete electrochemical environment.[21] For instance, it does not describe the effect of applied potential on activation barriers, the geometries of reacting species, nor the energies any non-PCET step.

In contrast, grand-canonical density functional theory (GC-DFT) provides a rigorous theoretical framework to explicitly account for the influence of the electrochemical potential and charged species present in the electrolyte on the free energies of reaction steps.[22,23] In GC-DFT, the electrostatic potential contributions from both the electrode and electrolyte are explicitly incorporated into the DFT calculations, allowing the electrode electronic charge and the corresponding induced electrolyte ionic charge to respond self-consistently to the applied potential. This enables modeling of electrochemical interface processes beyond PCETs and potential-dependent binding of reaction intermediates while including the effects of electrolyte solvation. By considering these additional effects, GC-DFT provides a more comprehensive and fundamental description of how the reaction energy landscape shifts with electrode potential. The database introduced here leverages the realistic treatment of the GC-DFT approach to evaluate electrochemical reactions across a wide range of operating conditions.

To make clear comparisons between various electrocatalysts under different operating conditions, we have constructed **BEAST** DB: the **B**eyond-DFT **E**lectrochemistry with **A**ccelerated and **S**olvated **T**echniques database. BEAST DB tabulates applied potential and continuum solvation effects computed using GC-DFT, thus providing quantities that are important for rationalizing computed adsorption energies as a function of potential (e.g. charges on molecule and active site atoms) and facilitating the comparison of calculations at equivalent calculation parameters. Ultimately, BEAST DB will also provide adsorption energies and electronic structure at a beyond-DFT level for the most promising catalysts and chemistries. Herein, we introduce BEAST DB and provide an overview of its contents and methods.

Throughout, we include example studies of specific systems to illustrate the capabilities afforded by BEAST DB. We compare the CHE with solvation and GC-DFT methods, investigating and contrasting their impacts on thermodynamic activity descriptors and reaction intermediate scaling relationships. Furthermore, we explore electronic structure variations under applied potential and their implications for predicted reaction energetics. We discuss GW and random-phase approximation (RPA) energy level calculations of select catalysts to assess how beyond-DFT exchange-correlation effects influence computed reaction energetics and our plans to extend these calculations to the most promising catalysts in the database. Finally, we discuss how a database of properties computed with high-level theories such as GW and RPA enables machine learning (ML)-based delta-learning to achieve beyond-DFT accuracy at the computational cost of standard DFT.

## 2. Database Overview

BEAST DB is designed to study the reaction pathways of several electrochemical redox processes by explicitly accounting for the impact of solvation and electrode potential.[23,24] The database contains individual GC-DFT calculations of adsorbates on catalyst surfaces as well as the adsorbate molecules and surfaces separately to facilitate binding energy calculations. BEAST DB currently compiles the results of > 21,000 calculations, spanning four unique material spaces and four distinct electrochemical reaction pathways (**Figure 1**). The database entries include single metal atom nitrogen doped graphene catalysts (MNC), flat and stepped pure metal surfaces, binary covalent alloys (BCA), and bimetallic single atom alloys (SAA). The database currently comprises adsorbates for the study of $CO_2$ reduction (CO2R), oxygen evolution/reduction (OER/ORR), hydrogen evolution/reduction (HER/HOR), and nitrogen reduction (NRR) pathways, although additional adsorbates relevant to the study of other reactions will be incorporated in subsequent releases. The coverage of reactions and their corresponding intermediates are shown in **Figure 1**, which shows that HER/HOR is the most common reaction in the dataset followed by NRR, OER/ORR, and CO2R.

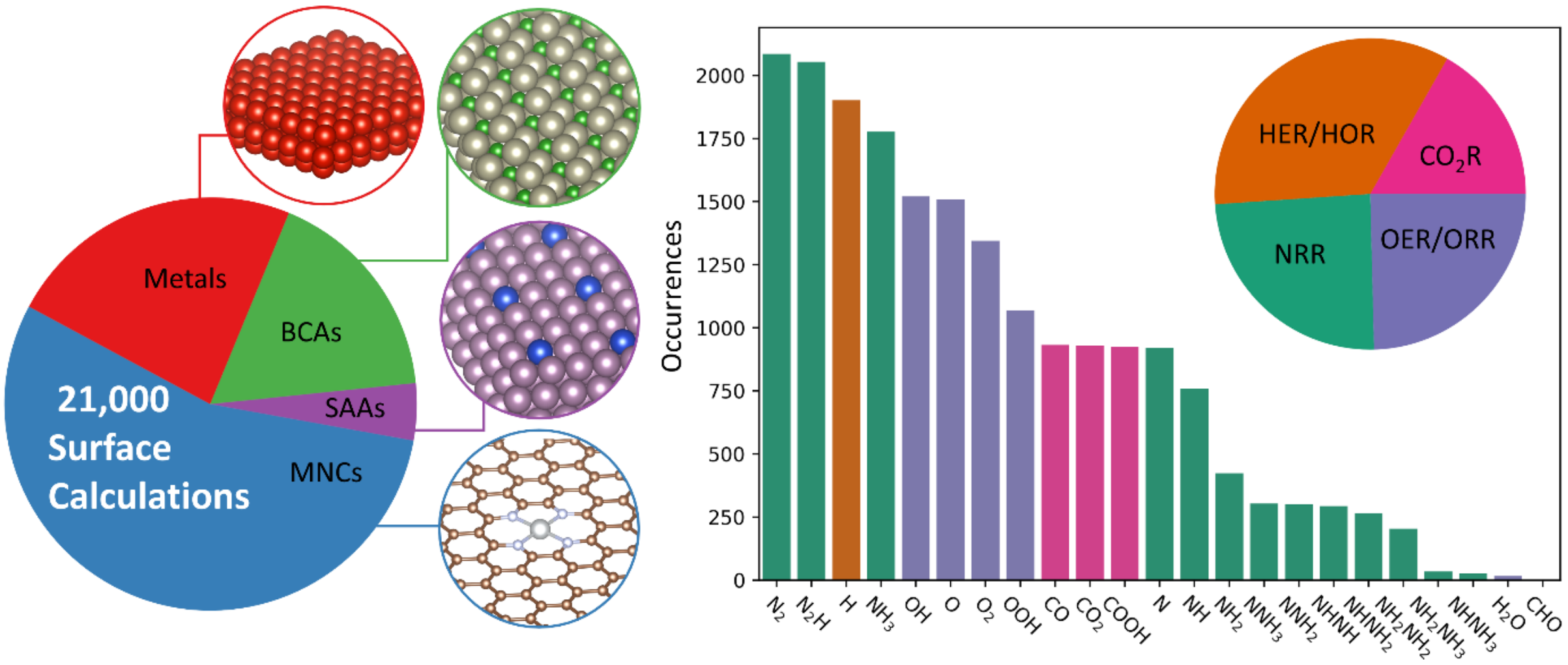

**Figure 1: Database Overview.** (left) The database is comprised of four main material classes: graphene-embedded single-atom catalysts (MNCs, 55%), monometallic flat and stepped metals (23%), binary covalent alloys (BCAs, 17%), and bimetallic single atom alloys (SAAs, 5%). (right) The distributions of reaction intermediate calculations are shown for the four main reactions; hydrogen evolution/reduction (HER/HOR), nitrogen reduction (NRR), oxygen evolution/reduction (OER/ORR), and $CO_2$ reduction (CO2R).

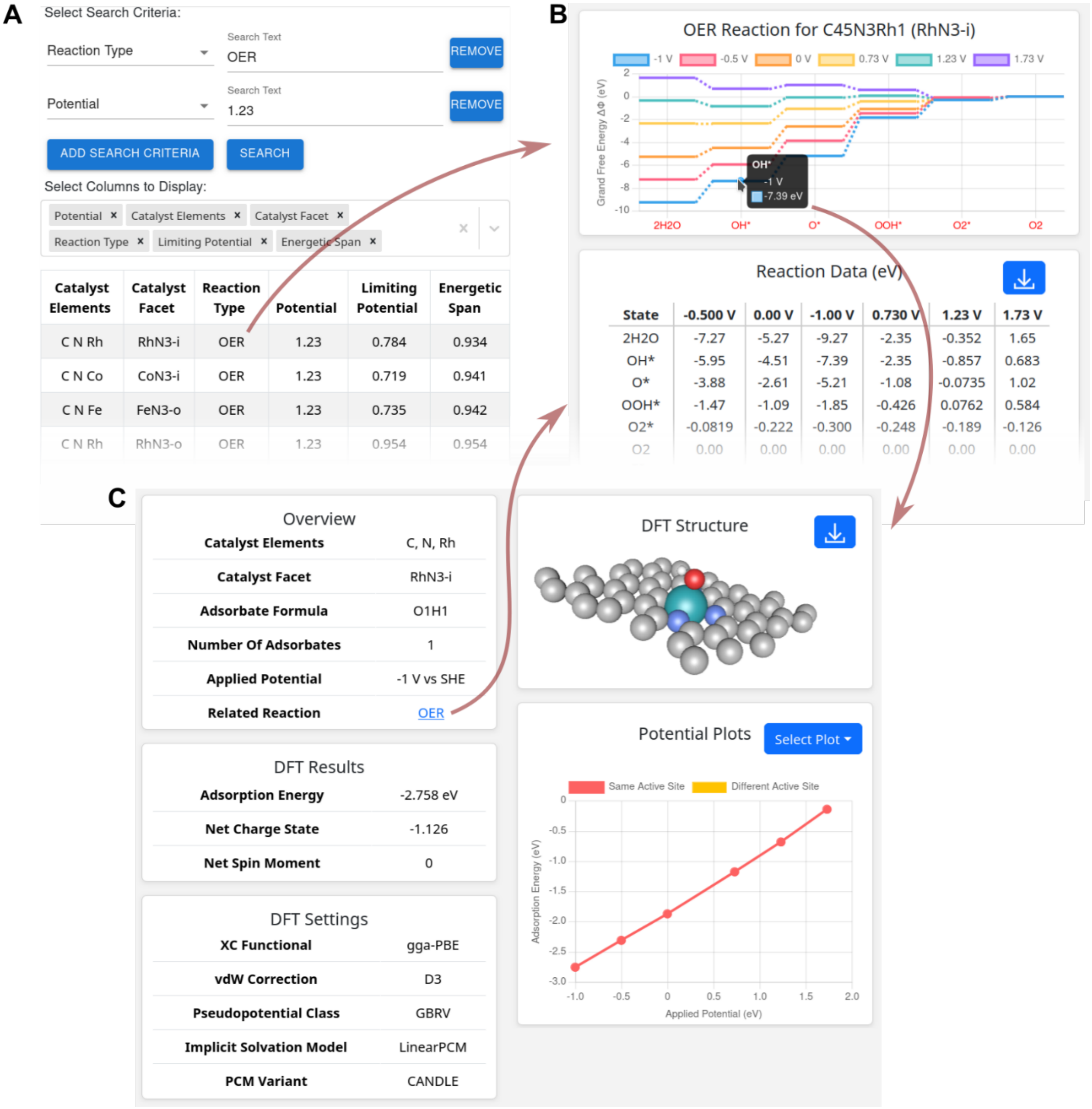

**Figure 2: User Interface**. (A) Reaction search based on catalyst elements, reaction type, potential and activity descriptors, shown here for OER at 1.23 V with sorting based on the energetic span. Each row links to (B) a detailed view of the reaction pathway as a function of potential. Each intermediate in the pathways links to (C) a detailed view of a specific catalyst with adsorbate calculation, including calculation details and plots of the structure and potential dependence of properties such as energy and charge (selectable). BEAST DB also allows searching directly for the adsorbates on catalysts (not shown here), linking to (C), which in turn links to corresponding reaction pathways.

The database interface allows users to search for reactions, catalysts, and adsorbates in several ways. **Figure 2** shows a typical workflow, starting from a search of catalysts by reaction type, with OER shown as an example (**Figure 2a**). This search can be filtered by any combination of catalyst elements, applied potential and thermodynamic descriptors such as limiting potential and energetic span, as well as sorted by any of these fields. The example shows the lowest energetic span OER catalysts at the selected potential. Each row links to a detailed reaction view (**Figure 2b**), showing the reaction pathways as a function of potential, along with a table of energies for all steps that can be downloaded for further external analysis. Each point on the pathway links to a detailed view of the catalyst + adsorbate configuration (**Figure 2c**), showing the structure, calculation details and resulting properties for that configuration. Importantly, this interface includes a potential-dependence plot that groups together all calculations available for a specific adsorbate on a surface, regardless of the electrode potential and adsorption site, *e.g.*, atop, bridge, hollow site, etc. This facilitates viewing the influence of the electrode potential as well as variations with respect to the adsorbed site on properties including adsorption energies, total charges, and atomic charges of the active site and adsorbate.[25] This also allows for understanding how charge transfer occurs during a reaction, how charge transfer changes as a function of potential, and how the observed charge transfer corresponds to observed trends in the reaction energetics as a function of potential. BEAST DB also allows searching directly for adsorbate configurations, leading to the interface in **Figure 2c**, which then links to all corresponding reaction pathways. Overall, the interface allows finding optimal catalysts and adsorbate configurations of interests and analyzing the impact of potential on all properties relevant for electrocatalytic reactions.

BEAST DB facilitates comparison of GC-DFT to the CHE applied to canonical DFT calculations. **Figure 3** compares the binding energies calculated by GC-DFT at a self-consistent bias vs. those calculated by canonical DFT under the same solvation. At 0 V vs. the standard hydrogen electrode (SHE), the agreement between CHE and GC-DFT is greatest, with a mean difference of 0.01 eV and a standard deviation of 0.74 eV. However, as the applied potential moves further from 0 V vs. SHE, so does the difference in calculated binding energies. At -1.00 and +1.23 V, the mean differences are -0.19 and 0.52 eV, respectively. We note that the adsorbates present at each of these biases are different, with calculations at +1.23 V composed of primarily OER/ORR species and calculations at -1.00 V composed of HER and NRR adsorbates. This difference in reaction intermediates may lead to the different directions in which the mean differences change, as it has been shown that some intermediate reaction steps display opposite sensitivities to applied potential.[26,27] These results show that, across the database, there are substantial differences in the CHE and GC-DFT calculated binding energies.

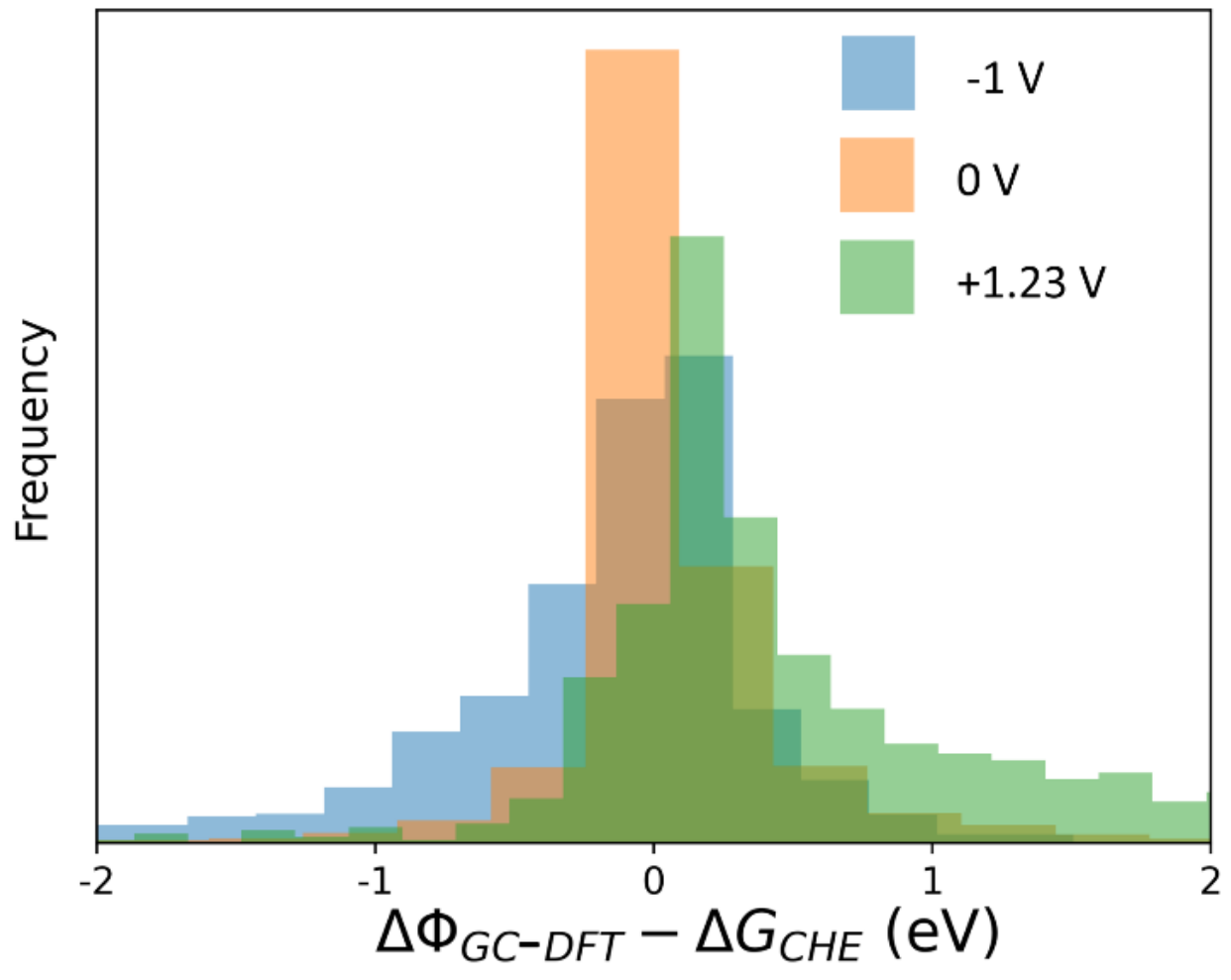

**Figure 3: Global CHE Comparison**. All voltages are with respect to SHE. Binding energies were calculated at self-consistent applied bias with GC-DFT from which binding energies calculated with CHE at neutral charge were subtracted. GC-DFT and CHE show better agreement the closer they are to 0 V vs SHE.

Although GC-DFT rigorously models the applied electrode potential, it is often still reliant on semilocal DFT functionals, which have been shown to be fundamentally limited in their accuracy towards predicting adsorption energies and relative favorability of binding sites, which is difficult to remedy without more advanced theory.[28] BEAST DB is primarily populated by results calculated using GC-DFT. However, to quantify errors associated with more computationally expedient approaches, we are also developing workflows that facilitate the significantly more expensive beyond-DFT GW and RPA calculations. In an upcoming release of BEAST DB, we will provide GW electronic structures and RPA total energies for a subset of the data consisting of transition metal surfaces (Ag, Cu, Ni, Pt, Ru) and a 2D catalyst ($FeN_4$@G) with small adsorbates important to the above electrochemical reactions (CO, $H_2$, $N_2$, $O_2$). Currently, these calculations are performed for surfaces in vacuum because methodological developments are still needed to facilitate the calculation of RPA total energies with solvation and applied potential.

# 3. Database Use Cases

## 3.1 Thermodynamic Activity Descriptors

Rapid evaluation and comparison of catalytic activity is important to aid researchers in selecting and designing performant catalysts. BEAST DB allows users to compare catalysts using multiple descriptors based on binding energies. The most widely used descriptor for electrocatalysts is the limiting potential ($U_L$),[17] which is defined as the minimum potential required to make all electrochemical steps thermoneutral or favorable. This descriptor is not only useful for ranking materials by predicted activity but also for estimating the potential at which the reaction proceeds. Our database facilitates screening of materials by limiting potential as shown in **Figure 4.**

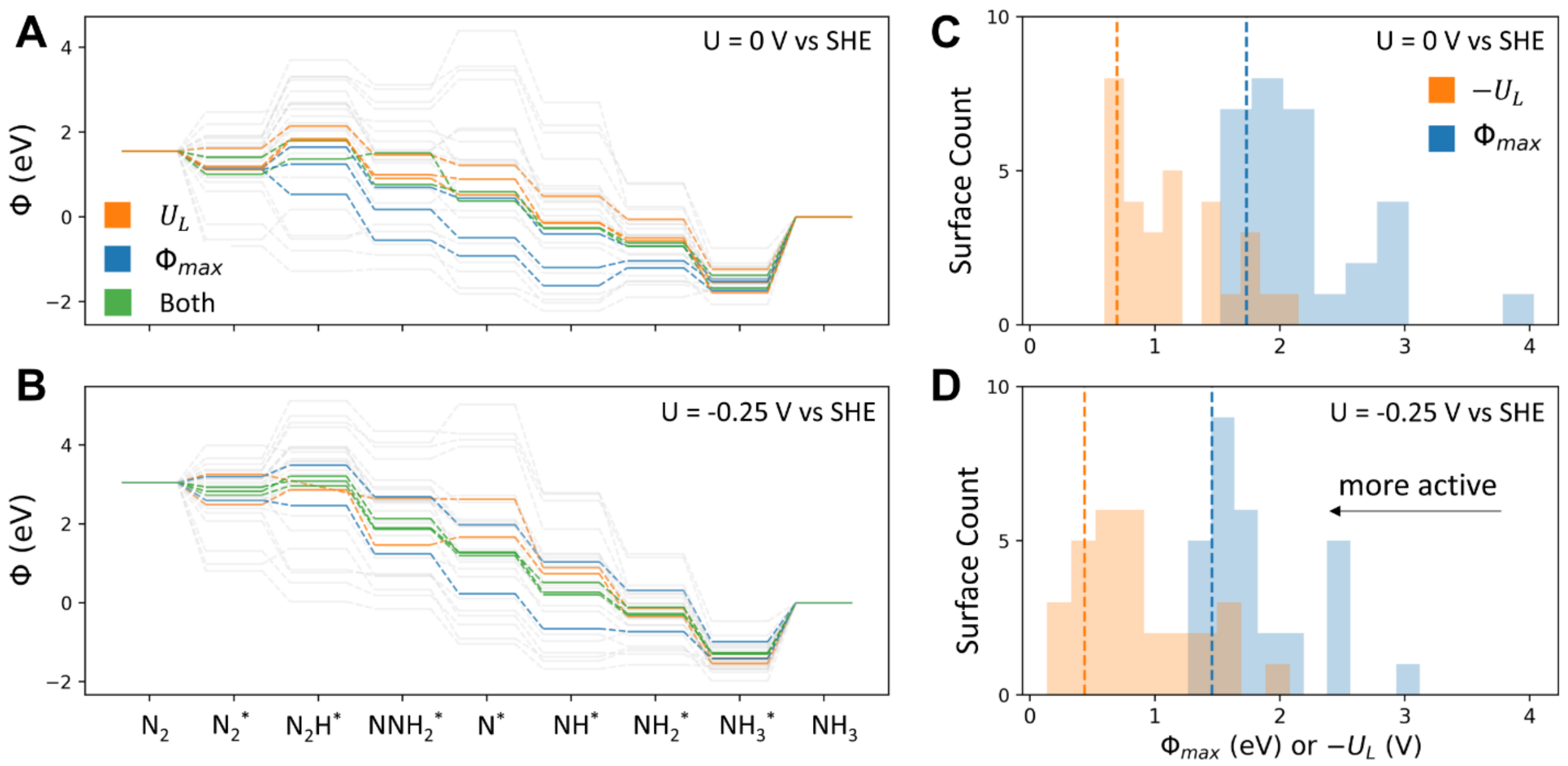

**Figure 4: Thermodynamic Descriptor Selection on Free Energy Pathways**. Free energy diagrams at 0 V and -0.25 V vs. SHE are shown in panels A and B, respectively. Panels C and D show $\Phi_{max}$ (blue) and -$U_L$ (orange) for all materials with a bin size of 0.25. Materials are sorted by the $\Phi_{max}$ and -$U_L$ descriptors and a cutoff is drawn for each descriptor at the 5$^{th}$ most active material according to each descriptor. The cutoffs are drawn in C and D with dashed lines. Materials that fall within the $\Phi_{max}$ cutoff are drawn in blue in figures A and B, and materials that fall within the -$U_L$ threshold are drawn in orange. Materials that fall within both descriptor thresholds, meaning both -$U_L$ and $\Phi_{max}$ predict these materials to be active, are drawn in green. -$U_L$ and $\Phi_{max}$ produce qualitative differences in catalyst ranking, with -$U_L$ selecting more weakly binding catalysts relative to $\Phi_{max}$ because $\Phi_{max}$ penalizes catalysts that under-bind $N_2^*$ and $N_2H^*$ after desorbing the over-bound $NH_3^*$ state.

The concept of limiting potential was initially defined for mechanisms comprised of purely PCET steps where the sensitivity of the step energy to potential is formally given by the change in the energy of the transferred electron.[17] However, GC-DFT facilitates the calculation of reaction steps that do not involve formal electron transfers (i.e., chemical steps), which are assumed to be insensitive to applied potential in the limiting potential scheme. The energetic span ($\Phi_{max}$)[29,30] allows for the evaluation of mechanisms that contain non-PCET steps that are still sensitive to potential (see **Section 3.3.1** for a potential-sensitive non-PCET step), which is calculated following the methods of our previous paper that used $\Phi_{max}$ to assess the NRR scaling relations.[31] **Figures 4A** and **4B** show the calculated free energy diagrams for the NRR on flat and stepped monometallic surfaces. Colored lines rank within the top five most active catalysts predicted by the -$U_L$ (orange) and $\Phi_{max}$ (blue) descriptor, while gray lines are all other less active catalysts screened. Panels **4C** and **4D** show the distribution of -$U_L$ and $\Phi_{max}$ across the monometallic dataset. Materials are sorted by their $U_L$ and $\Phi_{max}$values and the material with the fifth lowest $U_L$ and $\Phi_{max}$ defines a cutoff that separates the five most active materials predicted by either descriptor from the rest of the materials. Those cutoffs are drawn as dashed lines in **4C** and **4D** and their corresponding free energy pathways are colored by the descriptors that rank them in the top five most active catalysts. Using these two methods to select materials leads to mostly different sets with some overlap (green lines in **Figures 4A** and **4B**). $\Phi_{max}$ tends to select materials that bind adsorbates more strongly than the materials selected by $U_L$ because $\Phi_{max}$ penalizes materials that under-bind $N_2^*$ and $N_2H^*$ after desorbing the over-bound $NH_3^*$ state. In contrast, activity predictions based on $U_L$ are unaffected by the energetics of the $NH_3$ desorption step because it does not involve a formal PCET. Because $\Phi_{max}$ takes account of more elementary steps than $U_L$, its material ranking is expected to match experimental trends more closely. **Table 1** shows the thermodynamic activity descriptors and their limiting steps for the materials selected in **Figure 4** at both potentials. The material ranking changes between biases with both descriptors, which shows that the reaction step energies are not all perfectly linear with the applied potential. Furthermore, materials predicted to be active by $\Phi_{max}$ are not necessarily predicted to be active by $U_L$, although there is some overlap in both descriptors across potentials such as Co (111) and Ru (111).

**Table 1: Data for Materials from Figure 4**. $\Phi_{max}$ is in units of eV and $U_L$ in V vs. SHE. The six materials with the lowest $\Phi_{max}$ are shown and ordered by $\Phi_{max}$ at 0 V. The rank of each material represents the sorting order of the material according to the specified thermodynamic descriptor at the specified potential, with a rank of 1 denoting the predicted best catalyst.

| Material | 0 V vs SHE | | | | -0.25 V vs SHE | | | |
|---|---|---|---|---|---|---|---|---|
| | $\Phi_{max}$ | $U_L$ | $\Phi_{max}$ rank | $U_L$ rank | $\Phi_{max}$ | $U_L$ | $\Phi_{max}$ rank | $U_L$ rank |
| Fe (110) | 1.52 | -1.03 | 1 | 15 | 1.41 | -0.72 | 3 | 16 |
| Ru (111) | 1.56 | -0.70 | 2 | 6 | 1.27 | -0.31 | 1 | 3 |
| Mo (110) | 1.61 | -1.44 | 3 | 23 | 1.52 | -0.60 | 8 | 13 |
| Co (111) | 1.63 | -0.61 | 4 | 2 | 1.34 | -0.27 | 2 | 2 |
| Rh (211) | 1.68 | -0.69 | 5 | 5 | 1.59 | -0.56 | 12 | 11 |
| Re (111) | 1.73 | -1.20 | 6 | 20 | 1.46 | -0.74 | 6 | 17 |

Both the limiting potential and energetic span descriptors enable the comparison of electrocatalysts across a material space as a starting point for designing novel catalysts and selecting catalysts for experimental evaluation. However, we note that if the kinetically limiting steps are understood to be purely chemical, the energetic span descriptor is likely a better predictor of catalytic activity.[29,31]

### 3.2 Scaling Relations

Scaling relations describe the linear correlation between the binding energies of reaction intermediates on the catalyst surface and provide valuable insights into the catalytic activity of materials.[32,33] These relations give rise to the characteristic volcano relationships frequently observed between intermediate binding energies and catalytic activity. BEAST DB facilitates comprehensive scaling relations analysis across various applied electrode potentials using reaction mechanisms that go beyond PCET mechanisms, allowing for the identification of limiting steps and their sensitivity to the applied bias. To this end, scaling relations for a given reaction and material class can be calculated from BEAST DB. Then, individual materials can be mapped onto the resulting volcano plot to assess their proximity to the volcano peak and identify thermodynamic limitations.

Here, we highlight the scaling relations for the $CO_2$ reduction reaction on MNC catalysts. These scaling relations rely on a hypothetical mechanism, which follows the path: $CO_2 \rightarrow CO_2^* \rightarrow COOH^* \rightarrow CO^* \rightarrow CO$.[34] For $CO_2$R on MNC catalysts, the MNC with a Co metal center (Co-MNC) is found at the top of the volcano at 0 V vs. SHE, indicating its high predicted catalytic activity (**Figure 5**). Co-MNC undergoes the hypothesized mechanism, as shown in the figure. This computational prediction is consistent with experimental observations of MNC catalysts, which have also identified Co-MNC as one of the most active catalysts for $CO_2$R.[35] The position of Co-MNC near the yellow dashed scaling line suggests that its activity is limited by the CO desorption step, and selectively weakening the CO* binding energy could improve its performance.

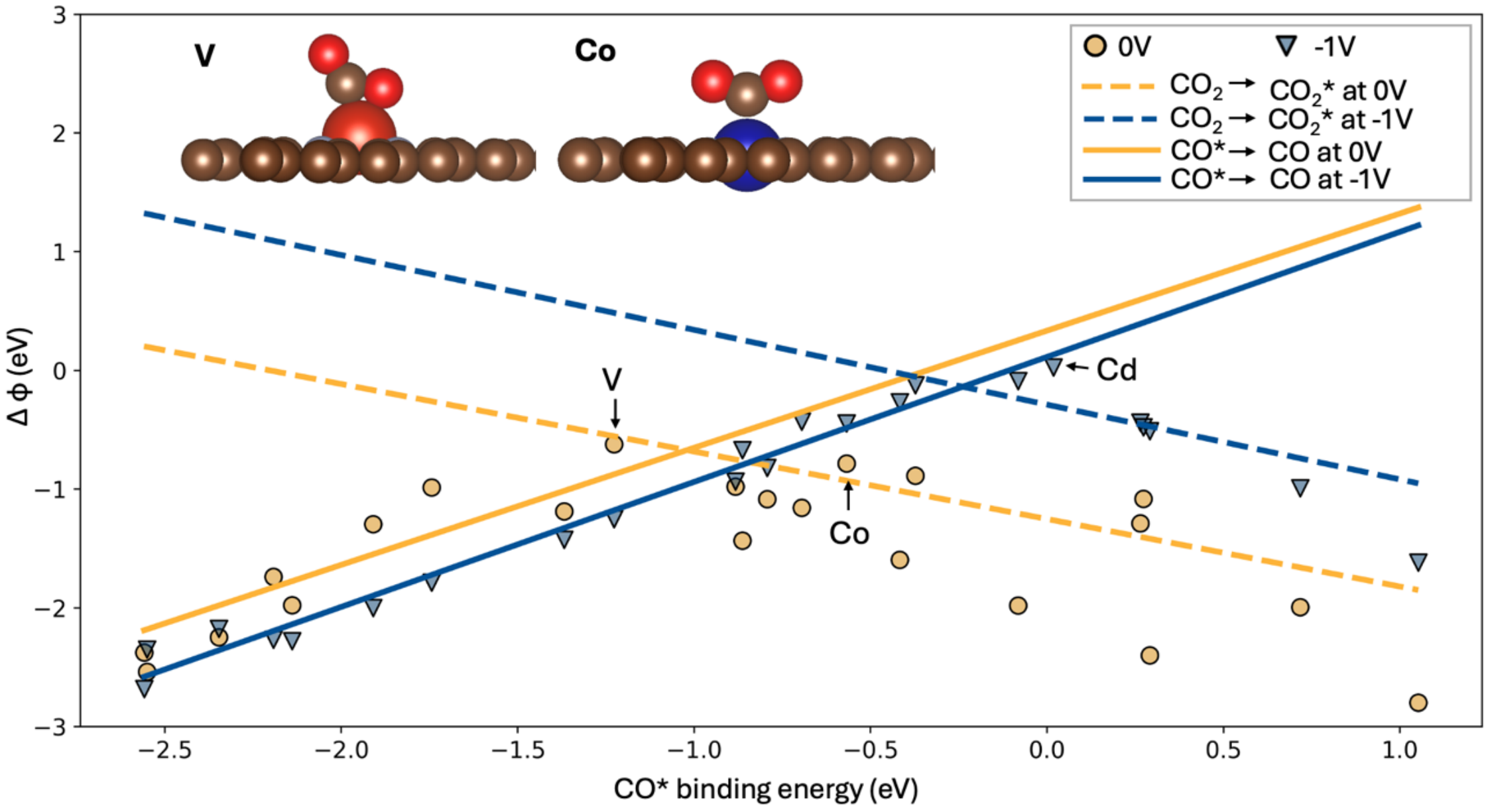

**Figure 5: Theoretical Volcano for CO2R on MN$_4$C:** Scaling of the largest elementary step change in grand free energy ($\Delta\phi$ of each catalyst with the binding energy of CO calculated at constant charge. The solid lines fit the CO desorption step to the constant charge CO binding energy, while the dashed lines fit the $CO_2^*$ adsorption step to the constant charge CO binding energy. Yellow and blue lines indicate 0 V vs. SHE and -1 V vs. SHE, respectively. The data points show the positions of various MNC catalysts on the volcano plot. Circles represent 0 V vs. SHE data and triangles represent -1 V vs. SHE data. The $CO_2$ binding geometries on vanadium and cobalt MNCs are shown on the upper left of the plot. The MNC with a Co metal center is located near the top of the volcano, indicating its high predicted catalytic activity at 0 V relative to other catalysts.

However, not all surfaces will undergo this exact mechanism. For instance, our GC-DFT calculations reveal that the Vanadium MNC (V-MNC) adsorbs $CO_2$ by bonding to both the carbon and one of the oxygens, suggesting that the scaling relations predicated on the assumed mechanism above are not predictive of the activity nor limiting steps for V-MNC. This is evident in **Figure 5**, where V-MNC lies above the scaling line at 0 V, and thus the scaling relations for our mechanism are not applicable. The figure excludes the MNCs that fail to adsorb $CO_2$. Moreover, Cd MNC is plotted for -1.00 V in **Figure 5** because GC-DFT predicts that $CO_2$ will not adsorb at 0 V vs. SHE but will adsorb at -1 V vs. SHE.

When the applied bias becomes more reducing (more negative), the peak of the volcano shifts upward and to the right (**Figure 5**). The vertical shift is expected because a more reducing potential makes the overall reaction more favorable. However, the vertical shift is relatively small because the $CO_2$ adsorption and CO desorption steps are chemical steps that are less sensitive to the applied potential compared to electrochemical steps that include formal electron transfers. The more pronounced shift to the right indicates that at a more reducing applied potential, the optimal catalyst should have a weaker affinity for CO.

We note that if the CHE model were used instead of GC-DFT, plotting these scaling relations at multiple potentials would not be possible, as the CHE model cannot differentiate between the limiting scaling lines at 0.00 and -1.00 V vs. SHE because they contain no formal electron transfers and the shift of the volcano peak at -1.00 V vs. SHE would not be observed.

## 3.3 Electronic Structure

### 3.3.1 GC-DFT Effects

The energetics of elementary electrochemical reactions are often well approximated by CHE, but steps that do not contain formal PCETs will not be impacted by the CHE correction. However, the energy change of such reactions is not necessarily independent of applied potential. One example is the adsorption of $CO_2$ to the catalyst surface, which is understood to be a reductive adsorption decoupled from any proton transfer.[34] Modeling this process with GC-DFT shows that the $CO_2$ adsorption step, which experimentalists have identified as the rate determining step,[36] is highly sensitive to applied bias. GC-DFT and the data in our database are appropriate tools for studying decoupled electron transfers but are also useful in understanding how the electronic structure changes with applied potential and how these changes manifest in reaction energetics.

For some surfaces reported in the dataset, the electronic structure of the adsorbate-surface complex changes substantially as a function of applied potential, which manifests as a bias dependence of steps that are classically thought of as purely chemical. For example, the capacitance of chemical bonds may allow for significant electron transfer as the potential changes,[37] as in the N-N bond in $N_2$ adsorbed onto TiC $(11\bar{1})$, as is shown in **Figure 6A**. Its adsorption energy decreases by 2.54 eV when the potential is lowered from +1.00 to -1.00 V vs. SHE. The $\pi^*$ antibonding states in $N_2$ (two orange peaks labeled with arrows in **Figure 6B**) lie partially below the Fermi level at +1.00 V vs. SHE and as potential is made more reducing, the antibonding orbitals are further populated, causing the N-N bond length to increase from 1.29 Å to 1.33 Å. This potential-dependent transfer of electrons in part leads to a highly potential-sensitive adsorption energy despite the lack of a formal electron transfer in the adsorption step. Population of these orbitals may signify a propensity to dissociate $N_2$, which could result in a step change in reaction mechanism as the applied potential changes.

Changes in electronic structure that deviate from the rigid-band approximation are important in understanding the impact of applied potential on reactivity.[37] Recently, several methods have been proposed to approximate the effects of the applied potential through *post-hoc* double layer charging and dipole-field interaction corrections.[38,39] While these approaches successfully describe electrochemical phenomena in many cases, they are incapable of capturing changes in electronic structure that are self-consistently calculated within GC-DFT and any ionic rearrangements they may induce.

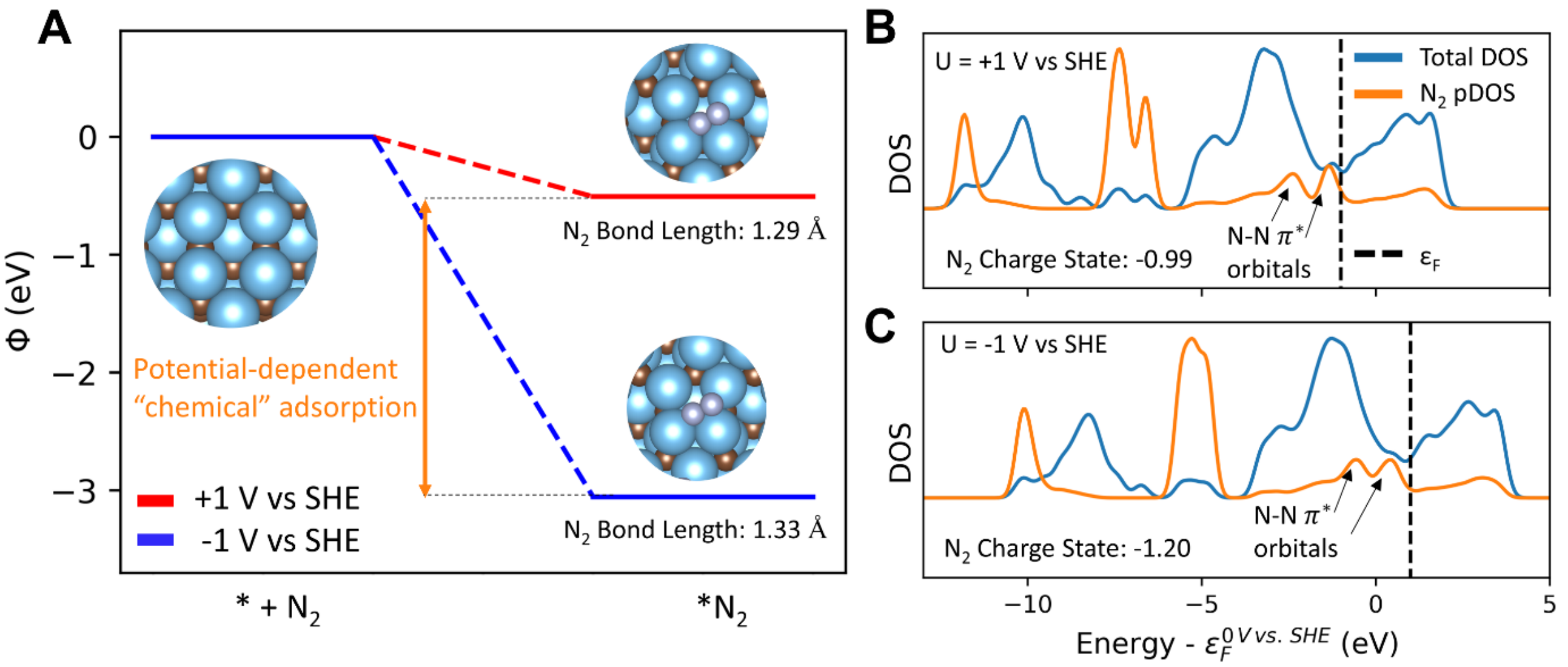

**Figure 6: Bias Dependent $N_2$ Adsorption:** (Left) adsorption of $N_2$ at 0 and -0.5 V vs. SHE on TiC (11$\bar{1}$). Although the step contains no formal electron transfer, it is highly sensitive to bias. (right) Total density of states (DOS) and projected density of states (pDOS) projected onto N atoms. The x-axis is zeroed at 0 V vs. SHE, showing that the Fermi level changes on an absolute scale with the electrode potentials. The net bond order decreases as indicated by the increased N-N bond length, greater population of $N_2$ $\pi^{*}$ orbitals, and increased charge localized on $N_2$.

### 3.3.2 Beyond-DFT Properties

In addition to solvation and potential effects, we are investigating how the beyond-DFT GW/RPA approaches impact electrocatalytic system predictions and potentially correct issues related to semilocal methods. For RPA total energies, we implemented algorithmic improvements that reduce the system-size scaling to improve RPA accessibility for catalytically relevant systems.[28,40] For GW electronic structures, we established a formalism to more seamlessly connect the GW framework to nonlocal implicit solvation models and directly incorporate the effect of liquid screening.[41] The development of a framework to calculate RPA total energies from solvated GC-DFT calculations using applied potential are the subject of ongoing work. Once complete, these features will be integrated into BEAST DB and released for public use. The next release of BEAST DB, however, will include RPA gas-phase corrections to the DFT adsorption energies for a selected set of catalysts.

We have used GW and RPA to screen small surface models and adsorbate combinations relevant to electrocatalysis (**Table S1**) and will be applying these techniques to larger surface models with more structurally diverse facets. The current set of GW electronic structures and RPA total energies includes CO, $H_2$, $N_2$, and $O_2$ adsorbed onto (1×1) Ag(100), Cu(100), Ni(100), Pt(100), Ru(001), and (3x3) $FeN_4$@G surfaces in vacuum at full coverage. These structures were generated from the same high-throughput DFT screening approach used to generate the structures for the rest of BEAST DB. **Figure 7A** shows that RPA destabilizes the adsorption energies of nearly all adsorbate/surface combinations relative to PBE calculations, with changes that are consistent with existing literature. We find that CO and $O_2$ adsorbed to $FeN_4$@G are particularly destabilized, while $N_2$ and $O_2$ adsorption to Ag(100) are particularly stabilized. The corresponding GW projected density of states (DOS) show that adsorption energy destabilization corresponds with the bonding orbitals of the adsorbates shifting to much lower energies relative to the Fermi level. An example of this for the Pt(100)+CO system is shown in **Figure 7B**. Although still in development, we believe that the incorporation of beyond-DFT methods will play a key role in further understanding the catalysts found from the above high-throughput screening approaches.

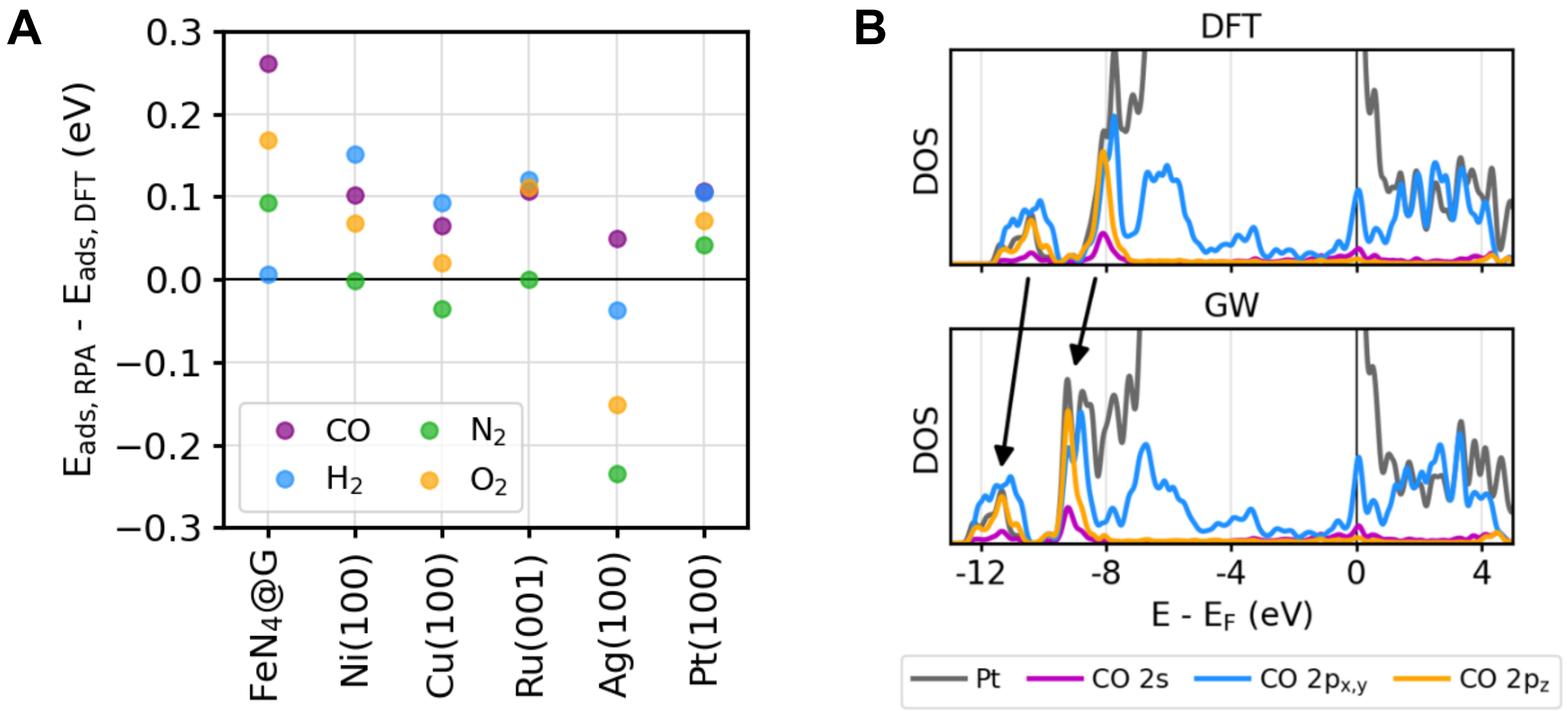

**Figure 7.** (A) Change in adsorption energy for CO, $H_2$, $N_2$, and $O_2$ adsorbed onto (1×1) Ag(100), Cu(100), Ni(100), Pt(100), Ru(001), and (3x3) $FeN_4$@G surfaces using RPA as compared to PBE. The molecules were adsorbed at full coverage on the transition metal surfaces. Positive values indicate destabilization using RPA. (B) GW projected density of states (DOS) for the Pt(100)+CO system as compared to the PBE DOS. The arrows highlight occupied bonding states that shift further in energy below the Fermi level using the GW approach.

### 3.4 Machine Learning Beyond-DFT Effects

Although GW quasiparticle eigenvalues have been shown to more accurately reflect the electronic structure of materials than DFT eigenvalues, the higher cost of GW calculations relative to DFT calculations has historically hindered such high-throughput calculations. Ongoing work in our project will leverage the BEAST DB to facilitate the development of ML models that Δ-learn GW quasiparticle eigenvalues from DFT eigenvalues. Efforts to create a large set of GW eigenvalues to Δ-learn is currently ongoing, however we have previously established a baseline framework for the capability by training an ML model that accurately predicts hybrid functional HSE06 eigenvalues of diverse bulk materials using a small feature set including DFT eigenvalues, projectors, and simple charge transfer metrics.[42] As an example that is relevant to the performance of Mg-air batteries,[43] **Figure 8** shows that the band structure for rock salt MgO predicted using the ML model significantly improves the semilocal calculated band energies relative to hybrid functionals. Our success in Δ-learning the HSE06 eigenvalues suggests that a similar approach could be used to predict GW eigenvalues. Furthermore, ongoing work is being performed to extend this approach to the prediction of surface with adsorbate electronic structures using a graph neural network, which accounts for local active site structure. We expect that this will further accelerate accurate generation and analysis of surface electronic structures, which will in turn be used to better understand trends in catalytic activity at the RPA level, as discussed in **Section 3.3.2**.

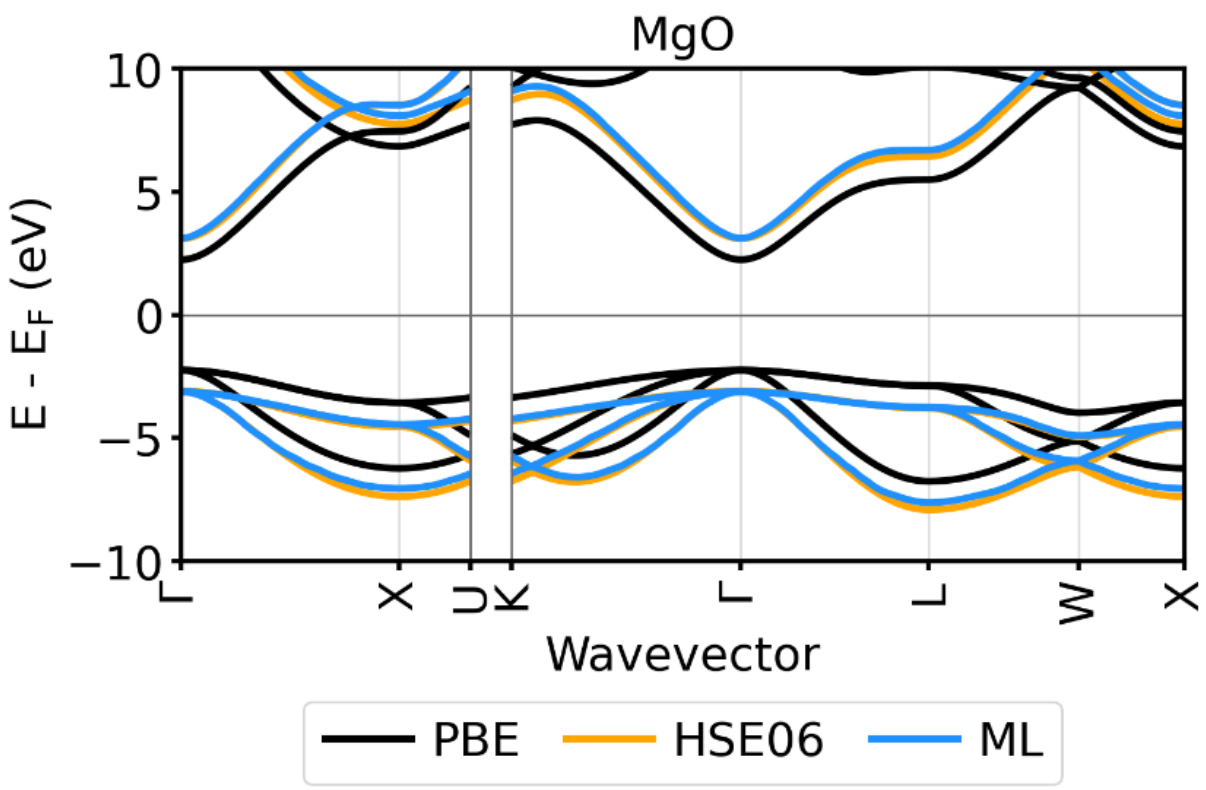

**Figure 8: Machine Learned Eigenvalues**. The rock salt MgO band structure predicted using a semilocal functional (PBE; black), a hybrid functional (HSE06; orange), and ML (blue).

## 4. Outlook

BEAST DB provides the community with a wealth of realistically computed electrochemical catalysis data and a set of tools to analyze the data. The database has continuing support and will be expanded significantly in the coming years to include more realistic surface models, more electrochemical reaction pathways, and calculations using higher-fidelity methods. The catalytic surfaces presented here are clean surfaces with low coverage of adsorbates (except the RPA calculated adsorption energies) with no competing species. In reality, catalytic surfaces are usually defected and have high coverage of various intermediates and electrolyte species, and the mechanisms on these surfaces can involve defects, *e.g.,* lattice oxygen. Furthermore, in addition to reaction thermodynamics, kinetics are crucial to providing insight into catalytic activity and selectivity. Current methods for calculating reaction rates suffer from non-robustness due to transition states being at a saddle point in contrast to an energy minimum. Additionally, both thermodynamics and kinetics are affected by the quality of the underlying theory used to describe the reaction energetics, and further work is needed to make beyond-DFT results more accessible to a larger range of reactions and chemistries. Finally, more advanced solvation models that capture atomic-scale liquid structure using classical-density functional theory or integral equation theories would allow for a seamless description of the electrocatalytic interface, including double layers and inhomogeneity of the fluid dielectric and ionic response from the interface to the bulk electrolyte. Ongoing work of the BEAST team targets generalizing these methods and robust enough to be applied to large sets of materials, on par with the current applicability of implicit solvation models.

With the current capabilities of BEAST DB and the planned work described above, the potential benefits to the catalysis community are significant. The ability to look at different catalysts and how their reaction thermodynamics and kinetics change as a function of potential and with different representations of surfaces coverages and defects will be beneficial to both theoretical and experimental catalysis researchers. Additionally, access to the atom-projected charges and spins at the catalyst active site and on the bound molecule using a consistent methodology facilitate comparison to experimental spectroscopic techniques such as XANES, and for the rationalization of observed trends. The combination of GW electronic structure and RPA reaction energetics with ML would enable an unprecedented level of accuracy in electrocatalytic calculations, with considerable benefits for the calculation of transition states and energetics in systems with localized electronic states. Access to a more complete description of electrocatalytic systems and more direct and transparent comparisons to experiments will enable the rational design of better electrocatalysts for a range of chemistries important to decarbonizing the world economy.

## 5. Methods

### 5.1 Density functional theory

The GC-DFT results in the database were calculated using the charge-asymmetric nonlocally determined local-electric (CANDLE) solvation model.[24] CANDLE has been shown to accurately account for solvation effects, accurately predict the pKa value of several molecules, better treat charge asymmetry between cations and anions, and allow for overall cell neutrality to be maintained when surfaces charge in GC-DFT calculations.

Within the grand canonical ensemble at a specified electrochemical potential vs. vacuum, $\mu$, the total grand free energy of a system, $\Phi[\mu]$, can be written as Eq. 1:

$$\Phi[\mu] = F_{DFT} - \mu N + E_{ZPE} + \int_0^T C_p \, dT - TS \tag{1}$$

where $F_{DFT}$ is the DFT Helmholtz energy and equal to the DFT total energy, $N$ is the total number of electrons in the system, $C_p$ is the system heat capacity at constant pressure, $T$ is the temperature, and S is the entropy. The DFT-computed Helmholtz energy is defined as $F_{DFT} = E_{DFT} - TS_{occ}$ for these systems because variable partial orbital occupations were used, where $S_{occ}$ is the entropy introduced due to the use of broadening, as is common for calculations involving metals. $E_{DFT}$ is the DFT internal energy. We note that the energies reported in BEAST DB do not include a concentration-dependent correction to the proton reference energy, thus effectively assuming a pH = 0, so that users can unambiguously add any desired correction their analysis.[17] The database reaction energies and adsorption energies only include free energy corrections for the molecular reference state energies under the assumption that vibrational corrections to the free energy for surface states cancel when calculating energy differences.

For the adsorption reaction shown below, where an adsorbate, $A^*$, adsorbates to an open surface site, $*$, to form the bound $A^*$ state, the grand canonical adsorption energy, $\Phi_{ads}[\mu]$, is thus calculated with Eq. 2:

$$A + \rightarrow A^*$$

$$\Phi_{ads}[\mu] = \Phi_{A^*}[\mu] - \left(\Phi_{Surf}[\mu] + \Phi_{MolcRef}[\mu]\right) \tag{2}$$

using the grand free energies of the adsorbed state, clean surface, and the appropriate set of molecular reference states. Note that although the potential must be kept constant across all grand free energies in Eq. 2, the number of electrons used to calculate each grand free energy may differ within the grand canonical ensemble. In addition to GC-DFT computed adsorption energies, BEAST DB contains adsorption energies calculated without applied potential using DFT Helmholtz free energies for uncharged systems. These adsorption energies facilitate comparisons to literature that has used the CHE approach. The electrochemical potential $\mu$ can be defined as an applied potential, $U$, relative to a particular reference electrode, $U_{Ref}$, using Eq. 3:

$$\mu = -\left(U_{Ref} + U\right) \tag{3}$$

Although the reference electrode potential may be defined using the widely used standard hydrogen electrode value of 4.44 eV, in this work we instead use $U_{Ref}$ = 4.66 eV, which is calibrated using the potentials of zero charge (PZC) of several transition metal surfaces for the CANDLE implicit solvation model.[24] Overall, these solvated, grand-canonical calculations capture the full, nonlinear variation of the grand free energies as a function of electrode potential, in contrast to the linear extrapolation of the grand free energy from neutral calculations in the conventional CHE approach.[17] We note that the above formalism was used to calculate the reaction pathways in BEAST DB as well. Additional details for these more complex calculations are described in detail at https://beastdb.nrel.gov/about.

## 5.2 GW approximation and random phase approximation

In this work and in the BEAST DB in the future, GW calculations are performed by solving the Dyson equation for the quasiparticle (QP) energies and wavefunctions (using Hartree atomic units),[44]

$$\left[\frac{-\nabla^2}{2} + V_{ion} + V_H + \Sigma\left(E_{\boldsymbol{nk}}^{\mathrm{QP}}\right)\right]\psi_{nk}^{\mathrm{QP}} = E_{\boldsymbol{nk}}^{\mathrm{QP}}\psi_{\boldsymbol{nk}}^{\mathrm{QP}} \tag{4}$$

where $\nabla$ is the kinetic energy operator, $V_{ion}$ is the ionic potential, $V_H$ is the Hartree potential, $\Sigma$ is the nonlocal, energy-dependent self energy, and $E_{\boldsymbol{nk}}^{\mathrm{QP}}$ and $\psi_{\boldsymbol{nk}}^{\mathrm{QP}}$ are the quasiparticle energies and wavefunctions, respectively. For both GW and RPA calculations, we use the single-particle wavefunctions, $\psi_{\boldsymbol{nk}}^{KS}$, from PBE DFT calculations as our quasiparticle wavefunctions (the diagonal approximation), along with the Kohn-Sham energies, $\varepsilon_{\boldsymbol{nk}}^{KS}$. Within the GW approximation, $\Sigma$ depends on the one-body Green's function, G, and the dynamically screened Coulomb interaction, W. W depends on the full-frequency dielectric matrix, which in turn depends on the independent-electron polarizability, $\chi^0$, as is standard[45] within the random phase approximation.[46] In the plane-wave basis, $\chi^0$ can be written as:

$$\chi_{\boldsymbol{GG'}}^0(\boldsymbol{q};\omega) = \sum_{\substack{n\,occ.\\ n'\,emp.}}\sum_{\boldsymbol{k}}\frac{2M_{nn'}^*(\boldsymbol{k},\boldsymbol{q},\boldsymbol{G})M_{nn'}(\boldsymbol{k},\boldsymbol{q},\boldsymbol{G'})}{\varepsilon_{n\boldsymbol{k}+\boldsymbol{q}} - \varepsilon_{n'\boldsymbol{k}} \pm \omega + i\delta} \tag{5}$$

with the matrix elements $M_{nn'}$ defined by

$$M_{nn'}(\boldsymbol{k},\boldsymbol{q},\boldsymbol{G}) \equiv \left\langle\psi_{n\boldsymbol{k}+\boldsymbol{q}}\middle|e^{i(\boldsymbol{q}+\boldsymbol{G})\cdot r}\middle|\psi_{n'\boldsymbol{k}}\right\rangle \tag{6}$$

Here, $\boldsymbol{q}$ is a vector in the first Brillouin zone, $\boldsymbol{G}$ and $\boldsymbol{G'}$ are reciprocal-lattice vectors, $\omega$ is frequency, $\delta$ is an infinitesimal that locates the poles of $\chi^0$ relative to the real frequency axis, and $n$ and $n'$ count occupied and empty bands, respectively. See Refs.[47–55] for detailed reviews of the GW and RPA methods.

The RPA correlation energy is computed from $\chi^0$ within the adiabatic connection-fluctuation dissipation theorem (ACFDT) as:

$$E_{corr}^{RPA} = \int_0^\infty \frac{d\omega}{2\pi} Tr\{ln[1 - v\chi^0(i\omega) + v\chi^0(i\omega)]\} \tag{7}$$

where Tr denotes the trace of a matrix and $v$ is the bare Coulomb interaction. RPA total energies were calculated by subtracting the exchange correlation contribution to the DFT Helmholtz energy, $E_{xc}$, from the DFT energy and adding in RPA correlation energy and the exact exchange energy computed with the input PBE orbitals.

### 5.3 High Throughput Computational Workflow

Bulk structures were optimized in JDFTx and surfaces were cut from the optimized structures using the pymatgen Python package surface generation functions. Surfaces were relaxed at constant potential and constant charge under the BEAST DB standard solvation parameters, which are described in detail at https://beastdb.nrel.gov/about. The pymatgen AdsorbateSiteFinder was used to enumerate unique active sites on the relaxed slabs. Adsorbate molecules were optimized in JDFTx under the standard solvation parameters before being placed on the adsorption sites of optimized surfaces. Adsorption calculations at a given potential were only initialized on surface structures that had been optimized at the same potential. Adsorbate structures were optimized, and binding energies were calculated from the optimized surface and adsorbate structures. The scripts used to run the calculations are hosted on our Github repository: https://github.com/Nick-Singstock/BEAST_DB_Manager. To standardize our workflows further and make external data contribution to BEAST DB possible, the next database milestone includes the refactoring of our workflows in FireWorks.[56]

## 6. Data Availability

All DFT data discussed in this study is available at https://beastdb.nrel.gov/. The results and methods used to perform the GW/RPA calculations in this study are presented in **Table S1** and accompanying text of the SI. The ML data presented in **Section 3.4** is available at https://github.com/Sutton-Research-Lab/ML_eigenshift.

## 7. Code Availability

All DFT calculations were run using the open-source code, JDFTx, available at https://jdftx.org/. Calculation workflows were managed using our high-throughput workflow manager available at https://github.com/Nick-Singstock/BEAST_DB_Manager. GW/RPA calculations were performed using the open-source BerkeleyGW software package available at: https://berkeleygw.org/download/.

## References

(1) Friend, C. M.; Xu, B. Heterogeneous Catalysis: A Central Science for a Sustainable Future. *Acc. Chem. Res.* **2017**, *50* (3), 517–521. https://doi.org/10.1021/acs.accounts.6b00510.

(2) Ehlers, J. C.; Feidenhans'l, A. A.; Therkildsen, K. T.; Larrazábal, G. O. Affordable Green Hydrogen from Alkaline Water Electrolysis: Key Research Needs from an Industrial Perspective. *ACS Energy Lett.* **2023**, *8* (3), 1502–1509. https://doi.org/10.1021/acsenergylett.2c02897.

(3) Ramachandran, R.; Chen, T.-W.; Veerakumar, P.; Anushya, G.; Chen, S.-M.; Kannan, R.; Mariyappan, V.; Chitra, S.; Ponmurugaraj, N.; Boominathan, M. Recent Development and Challenges in Fuel Cells and Water Electrolyzer Reactions: An Overview. *RSC Adv.* **2022**, *12* (43), 28227–28244. https://doi.org/10.1039/D2RA04853A.

(4) Resasco, J.; Bell, A. T. Electrocatalytic CO2 Reduction to Fuels: Progress and Opportunities. *Trends Chem.* **2020**, *2* (9), 825–836. https://doi.org/10.1016/j.trechm.2020.06.007.

(5) Qing, G.; Ghazfar, R.; Jackowski, S. T.; Habibzadeh, F.; Ashtiani, M. M.; Chen, C.-P.; Smith, M. R. I.; Hamann, T. W. Recent Advances and Challenges of Electrocatalytic N2 Reduction to Ammonia. *Chem. Rev.* **2020**, *120* (12), 5437–5516. https://doi.org/10.1021/acs.chemrev.9b00659.

(6) Resasco, J.; Abild-Pedersen, F.; Hahn, C.; Bao, Z.; Koper, M. T. M.; Jaramillo, T. F. Enhancing the Connection between Computation and Experiments in Electrocatalysis. *Nat. Catal.* **2022**, *5* (5), 374–381. https://doi.org/10.1038/s41929-022-00789-0.

(7) Mou, T.; Pillai, H. S.; Wang, S.; Wan, M.; Han, X.; Schweitzer, N. M.; Che, F.; Xin, H. Bridging the Complexity Gap in Computational Heterogeneous Catalysis with Machine Learning. *Nat. Catal.* **2023**, *6* (2), 122–136. https://doi.org/10.1038/s41929-023-00911-w.

(8) Sauer, J. The Future of Computational Catalysis. *J. Catal.* **2024**, *433*, 115482. https://doi.org/10.1016/j.jcat.2024.115482.

(9) Jain, A.; Ong, S. P.; Hautier, G.; Chen, W.; Richards, W. D.; Dacek, S.; Cholia, S.; Gunter, D.; Skinner, D.; Ceder, G.; Persson, K. A. Commentary: The Materials Project: A Materials Genome Approach to Accelerating Materials Innovation. *APL Mater.* **2013**, *1* (1), 011002. https://doi.org/10.1063/1.4812323.

(10) Ghiringhelli, L. M.; Carbogno, C.; Levchenko, S.; Mohamed, F.; Huhs, G.; Lüders, M.; Oliveira, M.; Scheffler, M. Towards Efficient Data Exchange and Sharing for Big-Data Driven Materials Science: Metadata and Data Formats. *Npj Comput. Mater.* **2017**, *3* (1), 1–9. https://doi.org/10.1038/s41524-017-0048-5.

(11) Saal, J. E.; Kirklin, S.; Aykol, M.; Meredig, B.; Wolverton, C. Materials Design and Discovery with High-Throughput Density Functional Theory: The Open Quantum Materials Database (OQMD). *JOM* **2013**, *65* (11), 1501–1509. https://doi.org/10.1007/s11837-013-0755-4.

(12) Curtarolo, S.; Setyawan, W.; Hart, G. L. W.; Jahnatek, M.; Chepulskii, R. V.; Taylor, R. H.; Wang, S.; Xue, J.; Yang, K.; Levy, O.; Mehl, M. J.; Stokes, H. T.; Demchenko, D. O.; Morgan, D. AFLOW:

An Automatic Framework for High-Throughput Materials Discovery. *Comput. Mater. Sci.* **2012**, *58*, 218–226. https://doi.org/10.1016/j.commatsci.2012.02.005.

(13) Pizzi, G.; Cepellotti, A.; Sabatini, R.; Marzari, N.; Kozinsky, B. AiiDA: Automated Interactive Infrastructure and Database for Computational Science. *Comput. Mater. Sci.* **2016**, *111*, 218–230. https://doi.org/10.1016/j.commatsci.2015.09.013.

(14) Chanussot, L.; Das, A.; Goyal, S.; Lavril, T.; Shuaibi, M.; Riviere, M.; Tran, K.; Heras-Domingo, J.; Ho, C.; Hu, W.; Palizhati, A.; Sriram, A.; Wood, B.; Yoon, J.; Parikh, D.; Zitnick, C. L.; Ulissi, Z. Open Catalyst 2020 (OC20) Dataset and Community Challenges. *ACS Catal.* **2021**, *11* (10), 6059–6072. https://doi.org/10.1021/acscatal.0c04525.

(15) *Catalyst Property Database*. https://cpd.chemcatbio.org/ (accessed 2024-05-24).

(16) Winther, K. T.; Hoffmann, M. J.; Boes, J. R.; Mamun, O.; Bajdich, M.; Bligaard, T. Catalysis-Hub.Org, an Open Electronic Structure Database for Surface Reactions. *Sci. Data* **2019**, *6* (1), 75. https://doi.org/10.1038/s41597-019-0081-y.

(17) Nørskov, J. K.; Rossmeisl, J.; Logadottir, A.; Lindqvist, L.; Kitchin, J. R.; Bligaard, T.; Jónsson, H. Origin of the Overpotential for Oxygen Reduction at a Fuel-Cell Cathode. *J. Phys. Chem. B* **2004**, *108* (46), 17886–17892. https://doi.org/10.1021/jp047349j.

(18) Studt, F. Grand Challenges in Computational Catalysis. *Front. Catal.* **2021**, *1*. https://doi.org/10.3389/fctls.2021.658965.

(19) Nørskov, J. K.; Bligaard, T.; Logadottir, A.; Kitchin, J. R.; Chen, J. G.; Pandelov, S.; Stimming, U. Trends in the Exchange Current for Hydrogen Evolution. *J. Electrochem. Soc.* **2005**, *152* (3), J23. https://doi.org/10.1149/1.1856988.

(20) Man, I. C.; Su, H.-Y.; Calle-Vallejo, F.; Hansen, H. A.; Martínez, J. I.; Inoglu, N. G.; Kitchin, J.; Jaramillo, T. F.; Nørskov, J. K.; Rossmeisl, J. Universality in Oxygen Evolution Electrocatalysis on Oxide Surfaces. *ChemCatChem* **2011**, *3* (7), 1159–1165. https://doi.org/10.1002/cctc.201000397.

(21) Oberhofer, H. Electrocatalysis Beyond the Computational Hydrogen Electrode. In *Handbook of Materials Modeling: Applications: Current and Emerging Materials*; Andreoni, W., Yip, S., Eds.; Springer International Publishing: Cham, 2018; pp 1–33. https://doi.org/10.1007/978-3-319-50257-1_9-1.

(22) Melander, M. M.; Kuisma, M. J.; Christensen, T. E. K.; Honkala, K. Grand-Canonical Approach to Density Functional Theory of Electrocatalytic Systems: Thermodynamics of Solid-Liquid Interfaces at Constant Ion and Electrode Potentials. *J. Chem. Phys.* **2018**, *150* (4), 041706. https://doi.org/10.1063/1.5047829.

(23) Sundararaman, R.; Goddard III, W. A.; Arias, T. A. Grand Canonical Electronic Density-Functional Theory: Algorithms and Applications to Electrochemistry. *J. Chem. Phys.* **2017**, *146* (11), 114104. https://doi.org/10.1063/1.4978411.

(24) Sundararaman, R.; Goddard, W. A. The Charge-Asymmetric Nonlocally Determined Local-Electric (CANDLE) Solvation Model. *J. Chem. Phys.* **2015**, *142* (6), 064107. https://doi.org/10.1063/1.4907731.

(25) Manz, T. A.; Limas, N. G. Introducing DDEC6 Atomic Population Analysis: Part 1. Charge Partitioning Theory and Methodology. *RSC Adv.* **2016**, *6* (53), 47771–47801. https://doi.org/10.1039/C6RA04656H.

(26) Alsunni, Y. A.; Alherz, A. W.; Musgrave, C. B. Electrocatalytic Reduction of CO2 to CO over Ag(110) and Cu(211) Modeled by Grand-Canonical Density Functional Theory. *J. Phys. Chem. C* **2021**, *125* (43), 23773–23783. https://doi.org/10.1021/acs.jpcc.1c07484.

(27) Tezak, C. R.; Gerits, S.; Rich, B.; Sutton, C.; Singstock, N.; Musgrave, C. B. Mapping the Binary Covalent Alloy Space to Pursue Superior Nitrogen Reduction Reaction Catalysts. *Adv. Energy Mater. n/a* (n/a), 2304559. https://doi.org/10.1002/aenm.202304559.

(28) Schimka, L.; Harl, J.; Stroppa, A.; Grüneis, A.; Marsman, M.; Mittendorfer, F.; Kresse, G. Accurate Surface and Adsorption Energies from Many-Body Perturbation Theory. *Nat. Mater.* **2010**, *9* (9), 741–744. https://doi.org/10.1038/nmat2806.

(29) Razzaq, S.; Exner, K. S. Materials Screening by the Descriptor Gmax(η): The Free-Energy Span Model in Electrocatalysis. *ACS Catal.* **2023**, *13* (3), 1740–1758. https://doi.org/10.1021/acscatal.2c03997.

(30) Kozuch, S.; Shaik, S. How to Conceptualize Catalytic Cycles? The Energetic Span Model. *Acc. Chem. Res.* **2011**, *44* (2), 101–110. https://doi.org/10.1021/ar1000956.

(31) Tezak, C. R.; Singstock, N. R.; Alherz, A. W.; Vigil-Fowler, D.; Sutton, C. A.; Sundararaman, R.; Musgrave, C. B. Revised Nitrogen Reduction Scaling Relations from Potential-Dependent Modeling of Chemical and Electrochemical Steps. *ACS Catal.* **2023**, 12894–12903. https://doi.org/10.1021/acscatal.3c01978.

(32) Bligaard, T.; Nørskov, J. K.; Dahl, S.; Matthiesen, J.; Christensen, C. H.; Sehested, J. The Brønsted–Evans–Polanyi Relation and the Volcano Curve in Heterogeneous Catalysis. *J. Catal.* **2004**, *224* (1), 206–217. https://doi.org/10.1016/j.jcat.2004.02.034.

(33) Abild-Pedersen, F.; Greeley, J.; Studt, F.; Rossmeisl, J.; Munter, T. R.; Moses, P. G.; Skúlason, E.; Bligaard, T.; Nørskov, J. K. Scaling Properties of Adsorption Energies for Hydrogen-Containing Molecules on Transition-Metal Surfaces. *Phys. Rev. Lett.* **2007**, *99* (1), 016105. https://doi.org/10.1103/PhysRevLett.99.016105.

(34) Brimley, P.; Almajed, H.; Alsunni, Y.; Alherz, A. W.; Bare, Z. J. L.; Smith, W. A.; Musgrave, C. B. Electrochemical CO2 Reduction over Metal-/Nitrogen-Doped Graphene Single-Atom Catalysts Modeled Using the Grand-Canonical Density Functional Theory. *ACS Catal.* **2022**, *12* (16), 10161–10171. https://doi.org/10.1021/acscatal.2c01832.

(35) Li, J.; Pršlja, P.; Shinagawa, T.; Martín Fernández, A. J.; Krumeich, F.; Artyushkova, K.; Atanassov, P.; Zitolo, A.; Zhou, Y.; García-Muelas, R.; López, N.; Pérez-Ramírez, J.; Jaouen, F. Volcano Trend in Electrocatalytic CO2 Reduction Activity over Atomically Dispersed Metal Sites on Nitrogen-Doped Carbon. *ACS Catal.* **2019**, *9* (11), 10426–10439. https://doi.org/10.1021/acscatal.9b02594.

(36) Deng, W.; Zhang, P.; Seger, B.; Gong, J. Unraveling the Rate-Limiting Step of Two-Electron Transfer Electrochemical Reduction of Carbon Dioxide. *Nat. Commun.* **2022**, *13* (1), 803. https://doi.org/10.1038/s41467-022-28436-z.

(37) Franco-Pérez, M.; Gázquez, J. L.; Ayers, P. W.; Vela, A. Revisiting the Definition of the Electronic Chemical Potential, Chemical Hardness, and Softness at Finite Temperatures. *J. Chem. Phys.* **2015**, *143* (15), 154103. https://doi.org/10.1063/1.4932539.

(38) Domínguez-Flores, F.; Melander, M. M. Approximating Constant Potential DFT with Canonical DFT and Electrostatic Corrections. *J. Chem. Phys.* **2023**, *158* (14), 144701. https://doi.org/10.1063/5.0138197.

(39) Agrawal, N.; Wong, A. J.-W.; Maheshwari, S.; Janik, M. J. An Efficient Approach to Compartmentalize Double Layer Effects on Kinetics of Interfacial Proton-Electron Transfer Reactions. *J. Catal.* **2024**, *430*, 115360. https://doi.org/10.1016/j.jcat.2024.115360.

(40) Del Ben, M.; da Jornada, F. H.; Antonius, G.; Rangel, T.; Louie, S. G.; Deslippe, J.; Canning, A. Static Subspace Approximation for the Evaluation of ${G}_{0}{W}_{0}$ Quasiparticle Energies within a Sum-over-Bands Approach. *Phys. Rev. B* **2019**, *99* (12), 125128. https://doi.org/10.1103/PhysRevB.99.125128.

(41) Clary, J. M.; Del Ben, M.; Sundararaman, R.; Vigil-Fowler, D. Impact of Solvation on the GW Quasiparticle Spectra of Molecules. *J. Appl. Phys.* **2023**, *134* (8), 085001. https://doi.org/10.1063/5.0160173.

(42) Adhikari, S.; Clary, J.; Sundararaman, R.; Musgrave, C.; Vigil-Fowler, D.; Sutton, C. Accurate Prediction of HSE06 Band Structures for a Diverse Set of Materials Using Δ-Learning. *Chem. Mater.* **2023**, *35* (20), 8397–8405. https://doi.org/10.1021/acs.chemmater.3c01131.

(43) Li, C.-S.; Sun, Y.; Gebert, F.; Chou, S.-L. Current Progress on Rechargeable Magnesium–Air Battery. *Adv. Energy Mater.* **2017**, *7* (24), 1700869. https://doi.org/10.1002/aenm.201700869.

(44) Hybertsen, M. S.; Louie, S. G. Electron Correlation in Semiconductors and Insulators: Band Gaps and Quasiparticle Energies. *Phys. Rev. B* **1986**, *34* (8), 5390–5413. https://doi.org/10.1103/PhysRevB.34.5390.

(45) Deslippe, J.; Samsonidze, G.; Strubbe, D. A.; Jain, M.; Cohen, M. L.; Louie, S. G. BerkeleyGW: A Massively Parallel Computer Package for the Calculation of the Quasiparticle and Optical Properties of Materials and Nanostructures. *Comput. Phys. Commun.* **2012**, *183* (6), 1269–1289. https://doi.org/10.1016/j.cpc.2011.12.006.

(46) Adler, S. L. Quantum Theory of the Dielectric Constant in Real Solids. *Phys. Rev.* **1962**, *126* (2), 413–420. https://doi.org/10.1103/PhysRev.126.413.

(47) Strinati, G.; Mattausch, H. J.; Hanke, W. Dynamical Aspects of Correlation Corrections in a Covalent Crystal. *Phys Rev B* **1982**, *25* (4), 2867–2888. https://doi.org/10.1103/PhysRevB.25.2867.

(48) Onida, G.; Reining, L.; Rubio, A. Electronic Excitations: Density-Functional versus Many-Body Green's-Function Approaches. *Rev Mod Phys* **2002**, *74* (2), 601–659. https://doi.org/10.1103/RevModPhys.74.601.

(49) Martin, P. C.; Schwinger, J. Theory of Many-Particle Systems. I. *Phys Rev* **1959**, *115* (6), 1342–1373. https://doi.org/10.1103/PhysRev.115.1342.

(50) Hedin, L.; Lundqvist, S. Effects of Electron-Electron and Electron-Phonon Interactions on the One-Electron States of Solids; Seitz, F., Turnbull, D., Ehrenreich, H., Eds.; Solid State Physics; Academic Press, 1970; Vol. 23, pp 1–181. https://doi.org/10.1016/S0081-1947(08)60615-3.

(51) Hedin, L. New Method for Calculating the One-Particle Green's Function with Application to the Electron-Gas Problem. *Phys Rev* **1965**, *139* (3A), A796–A823. https://doi.org/10.1103/PhysRev.139.A796.

(52) Godby, R. W.; Schlüter, M.; Sham, L. J. Self-Energy Operators and Exchange-Correlation Potentials in Semiconductors. *Phys Rev B* **1988**, *37* (17), 10159–10175. https://doi.org/10.1103/PhysRevB.37.10159.

(53) Farid, B.; Daling, R.; Lenstra, D.; van Haeringen, W. GW Approach to the Calculation of Electron Self-Energies in Semiconductors. *Phys Rev B* **1988**, *38* (11), 7530–7534. https://doi.org/10.1103/PhysRevB.38.7530.

(54) Aulbur, W. G.; Jönsson, L.; Wilkins, J. W. Quasiparticle Calculations in Solids; Ehrenreich, H., Spaepen, F., Eds.; Solid State Physics; Academic Press, 2000; Vol. 54, pp 1–218. https://doi.org/10.1016/S0081-1947(08)60248-9.

(55) Aryasetiawan, F.; Gunnarsson, O. The GW Method. *Rep. Prog. Phys.* **1998**, *61* (3), 237. https://doi.org/10.1088/0034-4885/61/3/002.

(56) Jain, A.; Ong, S. P.; Chen, W.; Medasani, B.; Qu, X.; Kocher, M.; Brafman, M.; Petretto, G.; Rignanese, G.-M.; Hautier, G.; Gunter, D.; Persson, K. A. FireWorks: A Dynamic Workflow System Designed for High-Throughput Applications. *Concurr. Comput. Pract. Exp.* **2015**, *27* (17), 5037–5059. https://doi.org/10.1002/cpe.3505.

## Acknowledgements

All authors gratefully acknowledge support from the U.S. Department of Energy, Office of Science, Basic Energy Sciences (DE-SC0022247). This work utilized computational resources from the National Energy Research Scientific Computing Center (NERSC) on the Perlmutter supercomputer.

## Competing Interests

The authors declare no competing interests.